\documentstyle[11pt,amssymb,epsf]{article}

\def\ben{\begin{equation}}
\def\een{\end{equation}}

\let\a=\alpha    
    
  \let\n=\nu

\let\C=\Chi

\def\nn{\nonumber} \def\bd{\begin{document}} \def\ed{\end{document}}
\def\ds{\documentstyle} \let\fr=\frac \let\bl=\bigl \let\br=\bigr
\let\Br=\Bigr \let\Bl=\Bigl
\let\bm=\bibitem
\let\na=\nabla
\let\pa=\partial \let\ov=\overline
\newcommand{\be}{\begin{equation}}
\newcommand{\ee}{\end{equation}}
\def\ba{\begin{array}}
\def\ea{\end{array}}
\def\ft#1#2{{\textstyle{{\scriptstyle #1}\over {\scriptstyle #2}}}}
\def\fft#1#2{{#1 \over #2}}
\def\del{\partial}
\def\vp{\varphi}
\def\sst#1{{\scriptscriptstyle #1}}
\def\oneone{\rlap 1\mkern4mu{\rm l}}
\def\td{\tilde}
\def\wtd{\widetilde}
\def\ie{\rm i.e.\ }
\def\dalemb#1#2{{\vbox{\hrule height .#2pt
        \hbox{\vrule width.#2pt height#1pt \kern#1pt
                \vrule width.#2pt}
        \hrule height.#2pt}}}
\def\square{\mathord{\dalemb{6.8}{7}\hbox{\hskip1pt}}}
\newcommand{\ho}[1]{$\, ^{#1}$}
\newcommand{\hoch}[1]{$\, ^{#1}$}
\newcommand{\bea}{\begin{eqnarray}}
\newcommand{\eea}{\end{eqnarray}}
\newcommand{\ra}{\rightarrow}
\newcommand{\lra}{\longrightarrow}
\newcommand{\Lra}{\Leftrightarrow}
\newcommand{\ap}{\alpha^\prime}
\newcommand{\bp}{\tilde \beta^\prime}
\newcommand{\tr}{{\rm tr} }
\newcommand{\Tr}{{\rm Tr} }
\def\0{{\sst{(0)}}}
\def\1{{\sst{(1)}}}
\def\2{{\sst{(2)}}}
\def\3{{\sst{(3)}}}
\def\4{{\sst{(4)}}}
\def\5{{\sst{(5)}}}
\def\6{{\sst{(6)}}}
\def\7{{\sst{(7)}}}
\def\8{{\sst{(8)}}}
\def\n{{\sst{(n)}}}
\def\cA{{{\cal A}}}
\def\cF{{{\cal F}}}
\def\tV{\widetilde V}
\def\tW{\widetilde W}
\def\tH{\widetilde H}
\def\tE{\widetilde E}
\def\tF{\widetilde F}
\def\tA{\widetilde A}
\def\im{{{\rm i}}}
\def\tY{{{\wtd Y}}}
\def\ep{{\epsilon}}
\def\vep{{\varepsilon}}
\def\R{\rlap{\rm I}\mkern3mu{\rm R}}
\def\bD{{{\bar D}}}

\def\R{\rlap{\rm I}\mkern3mu{\rm R}}
\def\bD{{{\bar D}}}
\def\R{{{\Bbb R}}}
\def\C{{{\Bbb C}}}
\def\H{{{\Bbb H}}}
\def\CP{{{\Bbb C}{\Bbb P}}}
\def\RP{{{\Bbb R}{\Bbb P}}}
\def\Z{{{\Bbb Z}}}
\def\bA{{{\Bbb A}}}
\def\bB{{{\Bbb B}}}
\def\bC{{{\Bbb C}}}
\def\bD{{{\Bbb D}}}
\def\bZ{{{\Bbb Z}}}
\def\Re{{{\frak{Re}}}}
\def\Im{{{\frak{Im}}}}
\def\cosec{{\,\hbox{cosec}\,}}

\newcommand{\tamphys}{\it Center for Theoretical Physics,
Texas A\&M University, College Station, TX 77843, USA}
\newcommand{\umich}{\it Michigan Center for Theoretical Physics,
University of Michigan\\ Ann Arbor, MI 48109, USA}
\newcommand{\upenn}{\it Department of Physics and Astronomy,
University of Pennsylvania\\ Philadelphia,  PA 19104, USA}
\newcommand{\SISSA}{\it  SISSA-ISAS and INFN, Sezione di Trieste\\
Via Beirut 2-4, I-34013, Trieste, Italy}

\newcommand{\ihp}{\it Institut Henri Poincar\'e\\
  11 rue Pierre et Marie Curie, F 75231 Paris Cedex 05}

\newcommand{\damtp}{\it DAMTP, Centre for Mathematical Sciences,
 Cambridge University\\ Wilberforce Road, Cambridge CB3 OWA, UK}
\newcommand{\itp}{\it Institute for Theoretical Physics, University of
California\\ Santa Barbara, CA 93106, USA}

\newcommand{\auth}{M. Cveti\v{c}\hoch{\dagger}, G.W. Gibbons\hoch{\sharp},
H. L\"u\hoch{\star} and C.N. Pope\hoch{\ddagger}}

\begin{document}
\begin{flushright}
\hfill{DAMTP-2001-30}\ \ \ {CTP TAMU-03/02}\ \ \ {UPR-955-T}\ \ \
{MCTP-02-13}\\
{March 2002}\ \ \
{hep-th/0203060}
\end{flushright}


\begin{center}
{ \large {\Large\bf Almost Special Holonomy in Type IIA\&M Theory}}

\vspace{20pt}
\auth

\vspace{3pt}
{\hoch{\dagger}\upenn}

\vspace{3pt}


\vspace{3pt}
{\hoch{\sharp}\damtp}

\vspace{3pt}
{\hoch{\star}\umich}

\vspace{3pt}
{\hoch{\ddagger}\tamphys}

\vspace{3pt}

\underline{ABSTRACT}
\end{center}

   We consider spaces $M_7$ and $M_8$ of $G_2$ holonomy and Spin(7)
holonomy in seven and eight dimensions, with a 
$U(1)$ isometry.  For metrics where the length of the associated
circle is everywhere finite and non-zero, one can perform a
Kaluza-Klein reduction of supersymmetric M-theory solutions
(Minkowski)$_4\times M_7$ or (Minkowski)$_3\times M_8$, to give
supersymmetric solutions (Minkowski)$_4\times Y_6$ or
(Minkowski)$_3\times Y_7$ in type IIA string theory with a
non-singular dilaton.  We study the associated six-dimensional and
seven-dimensional spaces $Y_6$ and $Y_7$ perturbatively in the regime
where the string coupling is weak but still non-zero, for which
the metrics remain Ricci-flat but that they no longer have special
holonomy, at the linearised level.  In fact they have ``almost special
holonomy,'' which for the case of $Y_6$ means almost K\"ahler,
together with a further condition.  For $Y_7$ we are led to introduce
the notion of an ``almost $G_2$ manifold,'' for which the associative
3-form is closed but not co-closed. We obtain explicit classes of 
non-singular metrics of almost special holonomy, associated with the 
near Gromov-Hausdorff limits of families of complete non-singular 
$G_2$ and Spin(7) metrics.

{\vfill\leftline{}\vfill
\vskip 5pt
\footnoterule
{\footnotesize \hoch{\dagger} Research supported in part by DOE grant
DE-FG02-95ER40893 and NATO grant 976951. \vskip -12pt} \vskip 14pt
{\footnotesize \hoch{\star} Research supported in full by DOE grant
DE-FG02-95ER40899 \vskip -12pt} \vskip 14pt
{\footnotesize  \hoch{\ddagger} Research supported in part by DOE
grant DE-FG03-95ER40917.\vskip  -12pt}}

\pagebreak
\setcounter{page}{1}

\tableofcontents
\addtocontents{toc}{\protect\setcounter{tocdepth}{3}}
\vfill\eject

\section{Introduction}

   Since the seminal work described in \cite{cahostwi}, the procedures
for obtaining four-dimensional physics from the compactification of
string theory have been extensively studied.  With the proposal of
an eleven-dimensional M-theory underlying string theory, it becomes of
interest to study the compactifications from the eleven-dimensional
standpoint, and to see what further lessons about the relations
between string theory and M-theory can be learned by this means.

   A key idea in the study of string-theory compactifications is that
the six-dimensional internal Calabi-Yau spaces can develop
singularities at limiting values of their modulus parameters, at which
additional massless four-dimensional states will occur.  Since the
moduli are themselves interpreted as four-dimensional massless scalar
fields, this means that the study of the four-dimensional low-energy
effective action requires a proper understanding of the regions in
modulus space where the singularities are approached.  It can be shown
that for the singularities of interest here, the Calabi-Yau metric
close to a generic singular point itself is of the form of the {\it
conifold}, which is the Ricci-flat metric
\be
ds_6^2 = dr^2 + r^2\, ds^2_{T^{1,1}}\label{conifold}
\ee
on the cone over the homogeneous Einstein space $T^{1,1}=(S^3\times
S^3)/U(1)_{\rm diag}$.  The metric is singular at the vertex of the
cone, at $r=0$.  As the moduli are moved slightly away from the
singular limit, the metric near to the previous conifold point is then
a smoothed-out version of (\ref{conifold}).  It can be shown that
there are two possible smoothed-out versions of (\ref{conifold}); one
is called the ``resolved conifold,'' with the conifold point blown up
to an $S^2$; the other is called the ``deformed conifold,'' with the
conifold point blown up to an $S^3$ \cite{candel}.  They are both
asymptotically conical (AC), approaching (\ref{conifold}) at large
distance.

   If Calabi-Yau compactifications are considered in the type IIA
string, then an immediate ``lift'' to M-theory compactifications can
be performed, simply by taking the direct product of the Calabi-Yau
manifold $Y_6$ and the M-theory circle $S^1$.  However, these are not
the only compactifications from $D=11$ that can be considered, and
more generally one can choose any compact Ricci-flat 7-manifold $M_7$.
In order to have supersymmetry, one requires that $M_7$ have special
holonomy, which in the irreducible case is the exceptional group
$G_2$.  Thus the $G_2$ manifold becomes natural compactification space
for the M-theory \cite{acharaya,atmava,atiwit}.  The direct product
$M_7=Y_6\times S^1$ is a degenerate example with $SU(3)\subset G_2$ holonomy.
More complicated possibilities, corresponding in the ten-dimensional
picture to turning on further fields including the Ramond-Ramond
vector and the dilaton in the type IIA theory, can also arise.

     Since an irreducible compact $G_2$ manifold $M_7$ cannot have any
continuous symmetries, and hence, in particular, no $U(1)$ action, it
follows that there cannot be an M-theory to type IIA reduction in
which exclusively massless fields (the Kaluza-Klein vector and the
dilaton) are excited.  However, if one focuses for now on the
structure of the metrics near to singular points, which for the
Calabi-Yau six-manifolds are approximated by the resolved and deformed
conifolds, then the seven-dimensional lifts of these non-compact
``building blocks'' can admit circle actions, and hence admit a clean
interpretation in terms exclusively of massless Kaluza-Klein field
excitations.

      Explicit examples of non-compact AC type of $G_2$ and spin(7)
manifolds were first construct in \cite{brysal,gibpagpop}.  The first
ALC manifold were constructed in \cite{cglpspin7}.  Recently, there
has been extensive work on constructing new non-compact $G_2$ and
Spin(7) manifolds \cite{konak,m3brane,kanyas1,brgogugu,%
cglpg2spin7,gukspa,cglporient,kanyas2,mconifold,brand,munify,hersfe}.
In some of this, it has been shown that there exist families of
non-compact $G_2$ metrics that are asymptotically locally conical
(ALC), which take the form of an $S^1$ bundle over an asymptotically
conical 6-metric, with the radius of the circle stabilising at large
distance.  The $G_2$ metrics have a non-trivial parameter (not merely
the overall scale) that allows the radius $R_0$ of the circle at
infinity to be adjusted while keeping some other measure of the
``scale size'' fixed.  At one extreme of the parameter one has
$R_0\longrightarrow 0$; this is the ``Gromov-Hausdorff limit'' in
which the metric approaches a direct product of a Calabi-Yau 6-metric
and the zero-radius $S^1$.  At the other extreme, the radius $R_0$
becomes infinite and then the $G_2$ metric is itself asymptotically
conical in a seven-dimensional sense.  Interpreting the $S^1$ as the
M-theory circle for the reduction to type IIA, we therefore have a
ten-dimensional Calabi-Yau reduction in the weak-coupling, or
Gromov-Hausdorff limit, whereas in the strong-coupling regime the
reduction is intrinsically non-perturbative and eleven-dimensional.  A
family of $G_2$ metrics denoted by $\bB_7$, whose weak-coupling limit
describes the deformed conifold, was considered in \cite{cglpg2spin7},
based on previous work in \cite{m3brane,brgogugu}.  An analogous
family, denoted by $\bD_7$, yielding the resolved conifold as a
weak-coupling limit was recently found in \cite{mconifold}.
Subsequently, a larger system of equations for $G_2$ metrics was
obtained in \cite{munify}, which encompasses both of the weak-coupling
limits.

    In this paper, we shall consider the relation between $G_2$
metrics admitting $U(1)$ actions and their associated circle
reductions in more detail.  In particular, we shall be concerned with
situations such as those described above where there is a non-trivial
parameter.  This allows one to probe the behaviour near to the
Gromov-Hausdorff limit, in which the string coupling is becoming weak
but is not yet zero.  Of particular interest are those cases where the
radius of the circle that stabilises to $R_0$ at infinity remains
non-zero (and finite) everywhere in the manifold, since in such cases
the dilaton resulting from the Kaluza-Klein reduction will be
everywhere finite.  The $\bD_7$ metrics found in \cite{mconifold} are
examples that exhibit this completely regular behaviour.  By contrast,
in the $\bB_7$ metrics studied in \cite{cglpg2spin7}, the length of the
circle goes to zero on the $S^3$ bolt in the core of the metric.  In
those cases with an everywhere-regular dilaton, one can study the
``near Gromov-Hausdorff'' regime of the seven-dimensional $G_2$ metric from a
six-dimensional Kaluza-Klein perspective, as a fully non-singular
perturbation away from the (Calabi-Yau) $\times S^1$ limiting case.  

    By this means, we are able to study non-singular perturbations of
the resolved conifold metric, which after lifting to $D=7$ become the
$\bD_7$ metrics with parameter $R_0$ close to the $R_0=0$
Gromov-Hausdorff limit.  In a similar vein, we also find non-singular
perturbations of Ricci-flat K\"ahler metrics on a complex line bundle
over $S^2\times S^2$.  These metrics, obtained for equal $S^2$ radii
in \cite{berber,pagpop}, and for unequal radii in \cite{pando2}, have
principal orbits $T^{1,1}/Z_2$.  After lifting the perturbed metrics
to seven dimensions, we obtain the ``near Gromov-Hausdorff regime'' of
another family of smooth $G_2$ metrics, denoted by $\wtd\bC_7$, which
are $\R^2$ bundles over $T^{1,1}$ \cite{munify}.  These too have the
feature that the radius of the circle action is everywhere finite and
non-zero.

    By studying the above perturbations of the original Ricci-flat K\"ahler 
six-dimensional spaces, we are able to determine what properties the
perturbed metrics should have in order that there exist a lift to give
$G_2$ holonomy in seven dimensions.  We find that at the level of
linearised perturbations away from the Calabi-Yau limit, the
six-dimensional metrics continue to be Ricci-flat, but they are no
longer K\"ahler.  In fact they still satisfy the almost-K\"ahler
condition, (namely that there is an almost complex structure and $dJ=0$), 
together with certain additional conditions that
are, nonetheless, weaker than the full K\"ahler condition.  (There is
a holomorphic 3-form whose real part is closed (in a suitable choice
of phase) but whose imaginary part is not.)  One
interesting aspect of this, therefore, is that at least for
non-compact spaces of the type we are considering here, one can have
smooth Ricci-flat perturbations that take the metric away from being
K\"ahler.

   After giving a general discussion of the perturbative scheme in
section 2, we then derive the details of the regular perturbations of
the resolved conifold and the complex line bundle over $S^2\times S^2$ in
section 3.

   In section 4 we consider a higher-dimensional analogue of the
previous discussion, in which eight-dimensional metrics of Spin(7)
holonomy, admitting circle actions, are reduced to seven dimensions.
In cases where there is a Gromov-Hausdorff limit, with the asymptotic
radius of the circle tending to zero, such a metric approaches a
seven-dimensional metric of $G_2$ holonomy times the zero-radius
circle.  We can again consider the regime in the vicinity of the
Gromov-Hausdorff limit, and examine the associated perturbations of
the seven-dimensional $G_2$ metric.  We find that here too, at the
linearised level the seven-dimensional metric remains Ricci-flat, but
it no longer has $G_2$-holonomy.  In fact the associative 3-form
$\Phi_\3$ that satisfied $d\Phi_\3=0=d{*\Phi_\3}$ in the unperturbed
$G_2$ limit still satisfies $d\Phi_\3=0$, but now
$d{*\Phi_\3}\ne0$.  By analogy with the case of K\"ahler manifolds we
propose to call such a seven-dimensional metric an ``almost $G_2$
metric.''

   A concrete example where we can obtain regular perturbations of a
$G_2$ metric is provided by the smooth asymptotically conical
metric $G_2$ metric with $SU(3)/(U(1)\times U(1))$ principal orbits.  
We obtain a class of linearised perturbations under which the metric
becomes ``almost $G_2$,'' while remaining Ricci-flat.  Upon lifting to
eight dimensions, this yields the near Gromov-Hausdorff limit of 
a class of smooth ALC Spin(7) metrics whose principal orbits are 
$N(1,-1)=SU(3)/U(1)_{(1,-1)}$, which were found in \cite{cglpg2spin7}.

   In section 5 we consider some further examples, in higher
dimensions, of Ricci-flat perturbations of Calabi-Yau metrics
that are almost K\"ahler but not K\"ahler.  Our starting points 
are the Ricci-flat K\"ahler metrics on complex line bundles over
$SO(n+2)/(SO(n)\times SO(2))$.   We show that one can obtain regular
perturbations in general, although now one no longer has a natural
interpretation involving a lift to one dimension higher.

\section{Kaluza-Klein reduction of $G_2$ metrics}\label{kksec}

\subsection{General discussion}\label{gendisc}

     In this section, we examine how a seven-dimensional metric of
$G_2$ holonomy that has a $U(1)$ isometry reduces to $D=6$.  In
subsequent sections we shall be interested in cases where the $G_2$
metric is asymptotically locally conical (ALC), locally approaching
the product of a circle of asymptotically constant radius (associated
with the above $U(1)$ action) and an asymptotically conical (AC)
six-dimensional metric.  Furthermore, our principal focus will be on
examples where the length of the circle remains non-vanishing (and
finite) everywhere in the $G_2$ manifold.  This additional assumption
that the $U(1)$ Killing vector has a bounded length is not, however,
required in the immediate discussion in this subsection.

  The seven-dimensional metric with a $U(1)$ isometry can be cast in
the adapted form of a Kaluza-Klein reduction ansatz.  In general, in a 
reduction from $(D+1)$ to $D$ dimensions, one has
\be
d\hat s_{D+1}^2 = e^{-2\a\, \phi}\, 
ds_D^2 + e^{2\a\, (D-2)\, \phi} (dz + \cA_\1)^2\,,
\label{kkred}
\ee
where $\a=[2(D-1)\, (D-2)]^{-1/2}$  
Using the tangent-space frame $\hat e^a =
e^{-\a\, \phi}\, e^a$, $\hat e^{\underline z}=e^{2\a\, (D-2)\, \phi}\, 
(dz+ \cA_\1)$, with $0\le a\le D-1$ and $e^a$ the vielbein for $ds_D^2$, 
the spin connection is given by
\bea
\hat \omega_{ab} &=& \omega_{ab} -\a\, 
e^{\a\, \phi}\, (\del_b\phi\, \hat e^a - 
 \del_a\phi\, \hat e^b) - \ft12 \cF_{ab}\, 
e^{\a\, D\, \phi}\, \hat e^{\underline z}\,,
\nn\\
\hat \omega_{a{\underline{z}}} 
&=& -\a\, (D-2)\, e^{\a\, \phi}\, 
\del_a\phi\, \hat e^{\underline z} 
- \ft12 \cF_{ab}\, e^{\a\, D\,\phi}\, 
\hat e^b\,.
\eea
The condition of Ricci-flatness in $(D+1)$ dimensions 
reduces to the equations of an Einstein-Maxwell-scalar system in $D$ 
dimensions, 
\bea
R_{ab} &=& \ft12\del_a\phi\, \del_b\phi +
\ft12 e^{2\a\, D\, \phi}\, (\cF^2_{ab} - 
\fft1{2(D-2)}\,  \cF_2^2\, g_{ab})\,,\nn\\
d{*d\phi} &=& \a\, D\, (-1)^{D-1}\, e^{2\a\, D\, \phi}\, 
{*\cF_\2}\wedge \cF_\2\,,\label{kkeom}\\
d(e^{2\a\, D\, \phi}\, {*\cF_\2}) &=&0\,,
\eea
which are described by the Lagrangian 
\be
{\cal L}_D = R\, {*\oneone}  -\ft12 {*d\phi}\wedge d\phi -\ft12
e^{2\a\, D\, \phi}\, {*\cF_\2}\wedge \cF_\2\,.
\ee

    The $G_2$ metric admits an ``associative 3-form''
$\hat \Phi_\3$, whose tangent-space components $\hat\Phi_{\sst{ABC}}$ give
the multiplication rules of the seven imaginary unit octonions 
$\iota_{\sst A}$:
\be
\iota_{\sst A}\, \iota_{\sst B}
 = -\delta_{\sst{AB}} + \hat\Phi_{\sst{ABC}}\, \iota_{\sst C}\,.
\ee
This implies that 
\be
\hat\Phi_{\sst{ABE}}\, \hat \Phi_{\sst{CDE}} = 
\delta_{\sst{AC}}\, \delta_{\sst{BD}} - 
\delta_{\sst{AD}}\, \delta_{\sst{BC}} + 
\ft16\, \ep_{\sst{ABCDE_1 E_2 E_3}}\, 
\hat\Phi_{\sst{E_1 E_2 E_3}}\,.\label{octmult}
\ee
The $G_2$ holonomy then follows from the conditions
\be
d\hat\Phi_\3 =0\,,\qquad d{\hat *\hat\Phi_\3} = 0\,,\label{clcocl}
\ee
where $\hat*$ denotes the seven-dimensional Hodge dual.  The conditions
(\ref{octmult}) and (\ref{clcocl}) imply that $\Phi_\3$ is 
covariantly constant \cite{joyce}.

   The dimensional reduction of $\hat\Phi_\3$ to $D=6$ is given by
\be
\hat \Phi_\3 = \Psi_\2\wedge (dz+\cA_\1) + \Psi_\3\,.
\ee
With the seven-dimensional indices decomposed as ${\sst A}=(a,6)$, where
$0\le a \le 5$, then in particular we can read off from (\ref{octmult}),
by setting two of the free indices to ``6,'' that
\be
\hat\Phi_{ab6}\, \hat\Phi_{bc6} = -\delta_{ab}\,,\label{acomp}
\ee
implying that $\hat J_{ab} \equiv \hat\Phi_{ab6}$ has the property of
being an almost complex structure in $D=6$:
\be
\hat J_{ab}\, \hat J_{bc} = -\delta_{ac}\,.
\ee
We can also then read off that
\be
\hat J_{ae}\, \hat \Phi_{bce}=\ft16 \ep_{abc d_1\, d_2\, d_3}\, 
\hat \Phi_{d_1 d_2 d_3}\,,\quad  
\hat\Phi_{abe}\, \hat\Phi_{cde} = \delta_{ac}\, \delta_{bd} - \delta_{ad}\, 
\delta_{bc} -\hat J_{ac}\, \hat J_{bd} + \hat J_{ad}\, \hat J_{bc}\,.
\ee
These conditions imply that $\hat\Phi_{abc}$ can be expressed as 
\be
\hat \Phi_{abc} = \Re(e^{\im\, \gamma}\, \ep_{abc})\,,
\ee
where $\gamma$ is an arbitrary phase angle and the antisymmetric
tensor $\ep_{abc}$ is holomorphic with respect to $\hat J_{ab}$.  

   The closure and co-closure of $\hat \Phi_\3$ imply 
\bea
&&d\Psi_\2 =0\,,\qquad d\Psi_\3 = - \Psi_\2\wedge d\cA_\1\,,\nn\\
&&d(e^{\fft2{\sqrt{10}}\phi} \, {*\Psi_\3}) =0\,,\qquad
d(e^{-\fft3{\sqrt{10}}\phi} \, {*\Psi_\2}) = e^{\fft2{\sqrt{10}}\phi}
{*\Psi_\3}\wedge d\cA_\1\,.\label{forms}
\eea

\subsection{Gromov-Hausdorff limit}

    We shall now specialise to the case where the $G_2$ metric is
ALC, with the $U(1)$ isometry acting on the circle of stabilised 
length at infinity.  

   It is worth remarking that in any Ricci-flat metric of
cohomogeneity one with an asymptotic region, in any dimension, the
length of any Killing vector $K$ increases monotonically with radius, and
thus it is bounded above by its length at infinity.  To see this, let
$f\equiv |K|^2$ be the squared length of $K$, which, of course,
satisfies $ \nabla^b\,\nabla_b\, K_a + R_{ab}\, K^b=0$.  Then we have
\be
\nabla^a\, \nabla_a\, f = 2 |\nabla_a\, K_b|^2 + 2 K^a\, \nabla^b\,
\nabla_b\, K_a =   2 |\nabla_a\, K_b|^2\,,\label{kvid}
\ee
the last step following from the fact that the metric is Ricci-flat.
Hence, integrating over $M(r_0)$, the manifold interior to radius
$r_0$, we have
\be
0\le \int_{M(r_0)} |\nabla_a\, K_b|^2\, \sqrt{g}\, d^nx = 
\int_{\del M(r_0)} \nabla_a\, f\, d\Sigma^a =\fft{df}{d
r}\Big|_{r=r_0}\, {\rm Vol}(\del M(r_0))\,,
\ee
which shows that $df/d r\ge0$, proving the result.  This implies that
the string coupling constant always decreases as one moves into the
interior, \ie towards the infra-red.  More generally, when the
Ricci-flat metric does not necessarily have cohomogeneity one,
(\ref{kvid}) implies that the quantity $f$ can have no interior
maximum.

    It is instructive first to consider the {\it Gromov-Hausdorff limit}, 
in which the radius of the circle at infinity is sent to zero, implying
that the Kaluza-Klein vector $\cA_\1$ also goes to zero, and hence from
(\ref{kkeom}) the dilaton $\phi$ becomes a constant. This can always
be achieved by using the trivial parameter in any Ricci-flat metric
that characterises the overall scale of the metric.  Typically, this
will lead to a conical singularity at the apex of the cone.  Of
greater interest, therefore, are cases where the ALC $G_2$ metric has 
a non-trivial parameter that allows one to send the radius of the
circle at infinity to zero while keeping the scale-size of the
bolt that resolves the apex of the cone non-vanishing.
In the Gromov-Hausdorff limit the seven-dimensional $G_2$ metric becomes
just a direct sum of a Ricci-flat six-metric and a circle,
\be
d\hat s_7^2 = dz^2 + ds_6^2\,,\label{ghlimit}
\ee
where the circle coordinate $z$ has been appropriately rescaled as the
limit is taken, so that it remains of non-vanishing period.  In the
cases where $d\hat s_7^2$ has a non-trivial parameter, the metric $ds_6^2$  
can be a non-trivial smooth AC 6-metric.\footnote{The four-dimensional
Taub-NUT metric is ALC, with only a trivial scale parameter.  
Nonetheless, in this case the Gromov-Hausdorff limit does give $S^1$
times a smooth 3-metric, namely $\R^3$ viewed as the cone over $S^2$.
The exceptional nature of this example arises because the cone metric
is itself non-singular here, on account of the base being a round 
sphere.}

   The special holonomy in $D=7$ requires that $ds_6^2$ must have
special holonomy; the most generic possibility is for the Ricci-flat
metric $ds_6^2$ to be K\"ahler, implying that it has $SU(3)$ holonomy.
Of course the Ricci-flat seven-metric in (\ref{ghlimit}) also then has
$SU(3)$ holonomy, and so it is a degenerate example of a $G_2$ metric.
The associative 3-form $\hat\Phi_\3$ in $D=7$ is then given
by
\be
\hat \Phi_\3 = \Psi_\2 \wedge dz + \Psi_\3\,,\label{3red}
\ee
where 
\be
\Psi_\2=J\,,\qquad  \Psi_\3=\Re(e^{\im\, \gamma}\, \ep_\3)\,,\label{gh23}
\ee
where $J$ is the K\"ahler form on $ds_6^2$, $\ep_\3$ is the 
closed holomorphic 3-form on $ds_6^2$ and $\gamma$ is an 
arbitrary constant phase angle.  It is easily seen that $\Psi_\2$ and
$\Psi_\3$ satisfy (\ref{forms}) with $\cA_\1=0$.  Conversely, we
may express the K\"ahler form and holomorphic 3-form in terms of 
$\Psi_\2$ and $\Psi_\3$ as 
\be
J=\Psi_\2\,,\qquad \ep_\3 = e^{-\im\, \gamma}\, (\Psi_\3 -\im\, 
{*\Psi_\3}) \,.\label{jpsi}
\ee

   One can of course reverse the argument; if one is
given a six-dimensional Ricci-flat K\"ahler metric $ds_6^2$, with 
K\"ahler form $J$ and holomorphic 3-form $\ep_\3$, then 
$d\hat s_7^2$ given in (\ref{ghlimit}) will be
a degenerate example of a seven-dimensional $G_2$ metric, with
holonomy $SU(3)\subset G_2$, and with associative 3-form given by 
(\ref{3red}) and (\ref{gh23}). 

\subsection{Linearised perturbation around the Gromov-Hausdorff limit}
\label{fosec}

   We may next consider a deformation away from the Gromov-Hausdorff
limit, in a linearised approximation, in which the Kaluza-Klein vector
$\cA_\1$ is assumed to be small, of order $\vep$, and only quantities
up to linear order in $\vep$ are retained.

  At this linearised level, it is evident from (\ref{kkeom}) that the
dilaton remains a constant, which without loss of generality we shall
take to be zero, and so the full set of equations governing the
six-dimensional fields become
\bea
&&R_{\mu\nu}=0\,,\qquad d\cF_\2 =0=d{*\cF_\2}\,,\nn\\
&&d\Psi_\2=0\,,\qquad
d{*\Psi_\2}-{*\Psi_\3}\wedge \cF_\2=0\,,\nn\\
&&d\Psi_\3+\Psi_\2\wedge \cF_\2=0\,,\qquad
d{*\Psi_\3}=0\,,\label{linsix}
\eea
plus terms of order $\vep^2$.  Note that although the six-dimensional
metric is still Ricci-flat in this linearised perturbation around the
Gromov-Hausdorff limit, it will no longer have special holonomy, since
$\Psi_\3$ is not closed and so we can no longer construct a
closed holomorphic 3-form $\ep_\3$ as in (\ref{jpsi}).

   Since the six-dimensional metric ceases to have special holonomy
once the linearised perturbation away from the Gromov-Hausdorff limit
is made, even though it remains Ricci-flat, this means that it will no
longer be K\"ahler.  It does however still remain, to order $\vep$,
{\it almost K\"ahler}, for which it is necessary only that there exist
an Hermitean almost-complex structure $J$ which, after lowering the
upper index to give a 2-form, satisfies the closure condition $dJ=0$.
(Co-closure is a consequence of these properties, since ${*J}=\pm
\ft12 J\wedge J$.)  The fact that the metric is no longer K\"ahler can
also be seen from the fact that $\Psi_\3$, which was given by
(\ref{gh23}) before the perturbation, now satisfies $d\Psi_\3 +
\Psi_2\wedge \cF_\2=0$, as given in (\ref{linsix}).\footnote{In
principle one might think that the metric could still be K\"ahler with
respect to a different choice of almost complex structure.  This
possibility can be excluded by considering an explicit example, and
verifying that $R_{abcd}\, \Gamma^{cd}$, the integrability condition
for a parallel spinor, has no zero eigenvalues once the perturbation
is turned on.  We have checked this explicitly for the Ricci-flat
deformations obtained in section \ref{expldisc}.} 
Note, however, that we shall still have $d{*\Psi_\3}=0$.

   The above discussion suggests that there can be a small
perturbations of certain Ricci-flat K\"ahler manifolds which remain
Ricci-flat, but where the metric is no longer K\"ahler, but rather is
almost K\"ahler, together with the further conditions discussed above.
Furthermore, we can then lift the the six-dimensional metric, by
making use of Kaluza-Klein ansatz (\ref{kkred}), and thereby obtain a
$G_2$ metric in seven dimensions.  An important point in following
this ``inverse'' procedure is that we can {\it derive} an expression
for the Kaluza-Klein vector that must be used for the lifting, using
only the already-known quantities in the perturbed six-dimensional
almost-K\"ahler metric.  To do this, we note from (\ref{linsix}) that
the almost-K\"ahler form $J=\Psi_\2$, and the 3-form $\Psi_\3$ obey
the relation $d\Psi_\3 + J\wedge \cF_\2=0$, where $\cF_\2=d\cA_\1$ and
$\cA_\1$ is the Kaluza-Klein vector.  It is straightforward to show
that in all complex dimensions greater than 2, if one has $J\wedge
\omega_\2=0$ where $\omega_\2$ is any 2-form and $J$ is constructed
from an almost complex structure, then it must be that $\omega_\2=0$.
Thus the equation
\be
d\Psi_\3 + J\wedge \cF_\2=0\label{cfdef}
\ee
can be solved uniquely for $\cF_\2$. If $d\cF_\2=0$ and $d{*\cF_\2}=0$, then 
writing $\cF_\2=d\cA_\1$ we have the Kaluza-Klein vector required for
lifting the metric to seven dimensions.

   To summarise, we have the following general result.  Starting from
a Ricci-flat K\"ahler metric $ds_6^2$, we make a linearised
perturbation that preserves the Ricci-flatness, while taking the
metric from K\"ahler to almost-K\"ahler.  If this perturbed metric
satisfies two further conditions, namely that for some choice of phase
angle $\gamma$ we have $d\Re(e^{\im\, \gamma}\, \ep_\3)=0$, and that
$\cF_\2$ defined by $d\Im(e^{\im\, \gamma}\, \ep_\3) + J\wedge \cF_\2=0$
is closed, then we can lift it to give a $G_2$ metric in $D=7$.  Note
that by a theorem of Sekigawa \cite{sekigawa}, any complete,
non-singular, compact, Ricci-flat, almost K\"ahler manifold must in
fact be K\"ahler.  It is not difficult to see, using identities 
derived in \cite{apodramor}, that this result generalises to the
non-compact cases we are considering here, as long as we insist on
non-singularity of the metric.  This means that at higher order, the
Ricci-flat almost K\"ahler conditions cease to hold.  (This accords
with the discussion in section \ref{gendisc}, since the dimensional
reduction of the smooth $G_2$ metrics away from the Gromov-Hausdorff
limit will obey (\ref{kkeom}) and (\ref{forms}), with $R_{ab}$
becoming non-zero beyond the linearised level.)

\section{Ricci-flat perturbations of Calabi-Yau 6-metrics}
\label{d6kah}

   In this section we shall study linearised perturbations around
certain classes of Ricci-flat K\"ahler 6-metrics, and make use of the
previous discussion to relate them to $G_2$ metrics in seven
dimensions.  Specifically, we shall consider perturbations of the
metric on the resolved conifold, with principal orbits
$T^{1,1}=(S^3\times S^3)/U(1)$, and of the smooth metrics on a complex
line bundle over $S^2\times S^2$, with principal orbits $T^{1,1}/Z_2$.
In the process, we shall obtain at the linearised level classes of
smooth Ricci-flat almost-K\"ahler metrics.  After lifting to $D=7$,
they turn out to correspond to previously-encountered smooth $G_2$ metrics
near to their Gromov-Hausdorff limits.  In fact since the existence 
of the $G_2$ metrics has until now been established by numerical 
integration, our new results provide additional analytic evidence for
their regularity.

\subsection{Metric ansatz}

   Our starting point is the class of cohomogeneity one metrics with
$SU(2)\times SU(2)\times U(1)$ isometries, where the principal orbits
are $T^{1,1}$ or $T^{1,1}/Z_2$.  These metrics encompass the deformed
and resolved conifolds \cite{candel}; and the complex line bundle over
$S^2\times S^2$ \cite{berber,pagpop} and its generalisation
\cite{pando2}.  (The generalisation allows the radii of the two $S^2$
factors to be specified independently, while they are equal in the
original example in \cite{berber,pagpop}.)  The metric ansatz will be
taken to be
\bea
ds_6^2 &=& dt^2 + a^2\, [(\Sigma_1 + g\, \sigma_1)^2 +
                          (\Sigma_2 + g\, \sigma_2)^2]
+ b^2\,  [(\Sigma_1 - g\, \sigma_1)^2 +
                          (\Sigma_2 - g\, \sigma_2)^2]\nn\\
&&+ c^2\, (\Sigma_3-\sigma_3)^2\,.\label{d6ans}
\eea
where $a$, $b$, $c$ and $g$ are functions only of the radial variable
$t$, and $\sigma_i$ and $\Sigma_i$ are left-invariant 1-forms of
$SU(2)\times SU(2)$.  They can be expressed in terms of Euler angles
as\footnote{Note that although there are ostensibly six coordinates
here, when one substitutes into (\ref{d6ans}) $\psi$ and $\wtd\psi$
appear only through the combination $\psi-\wtd\psi$.  The
parameterisation is a useful one for what follows later when we lift
to seven dimensions, where the remaining combination $\psi+\wtd\psi$
will form the fibre coordinate.}
\bea
&&\sigma_1+\im\, \sigma_2= e^{-\im\, \psi}\, (d\theta + \im\,
\sin\theta\, d\phi)\,,\qquad \sigma_3=d\psi + \cos\theta\,
d\phi\,,\nn\\
&&\Sigma_1+\im\, \Sigma_2= e^{-\im\, \wtd\psi}\, (d\td\theta + \im\,
\sin\td\theta\, d\td\phi)\,,\qquad \Sigma_3=d\wtd\psi + \cos\td\theta\,
d\td\phi\,,
\eea
and they satisfy
\be
d\sigma_i=-\ft12 \ep_{ijk}\, \sigma_j\wedge \sigma_k\,,\qquad
d\Sigma_i=-\ft12 \ep_{ijk}\, \Sigma_j\wedge \Sigma_k\,.
\ee
We shall take the vielbein basis to be 
\bea
&&e^0=dt\,,\quad e^1=a\, (\Sigma_1 + g\,
\sigma_1)\,,\quad e^2=a\, (\Sigma_2 + g\,
\sigma_2)\,,\quad e^3 =c\, (\Sigma_3 - \sigma_3)\,,\nn\\
&&e^4 = b\, (\Sigma_1 - g\, \sigma_1)\,,\quad
e^5 = b\, (\Sigma_2 - g\, \sigma_2)\,.\label{6bein}
\eea

   An invariant choice for an almost complex structure is 
\be
J= e^0\wedge e^3 + e^1\wedge e^5 - e^2\wedge e^4\,,\label{akj}
\ee
from which it follows that the following can be chosen as a holomorphic
complex vielbein:
\be
\ep^0= e^0 + \im\, e^3\,,\qquad \ep^1=e^1 + \im\, e^5\,,\qquad 
\ep^2= e^2 -\im\, e^4\,,
\ee
in terms of which we have $J=\fft{\im}{2}\, (\ep^0\wedge \bar\ep^0+
\ep^1\wedge \bar\ep^1+ \ep^2\wedge \bar\ep^2)$.  We then define the 
holomorphic 3-form
\be
\ep_\3 \equiv  \ep^0\wedge \ep^1\wedge \ep^2 =\ep_\3^R + \im\, \ep_\3^I\,,
\ee
with the real and imaginary parts given by 
\bea
\ep_\3^R &=& e^0\wedge e^1\wedge e^2 - e^0\wedge e^4\wedge e^5
            + e^3\wedge e^1\wedge e^4 + e^3\wedge e^2\wedge e^5\,,\nn\\
\ep_\3^I &=& e^3\wedge e^1\wedge e^2 - e^3\wedge e^4\wedge e^5
            - e^0\wedge e^1\wedge e^4 - e^0\wedge e^2\wedge e^5\,.
\eea
Note that since ${*\ep_\3}=\im\, \ep_\3$, we shall have
${*\ep_\3^R}=-\ep_\3^I$ and ${*\ep_\3^I} = \ep_\3^R$.

\subsection{Conditions for $SU(3)$ holonomy}

   Imposing the closure of $J$, $\ep_\3^R$ and $\ep_\3^I$ gives
\bea
dJ=0:&& 2 (a\, b){}^\cdot_{\phantom \Xi} 
+ c=0\,,\qquad 2(a\, b\, g^2){}^\cdot_{\phantom \Xi}  + c=0\,,
\label{djcon}\\
d\ep_\3^R=0:&& 2(a\, b\, c\, g){}^\cdot_{\phantom \Xi}  
+ (a^2+b^2)\, g=0\,,\label{deprcon}\\
d\ep_\3^I =0:&& [(a^2-b^2)\, c]{}^\cdot_{\phantom \Xi} =0\,,
\qquad [(a^2-b^2)\, c\, g^2]{}^\cdot_{\phantom \Xi}  =0\,,\nn\\
&&[(a^2+b^2)\, c\, g]{}^\cdot_{\phantom \Xi}  + 2a\, b\, g=0\,,\qquad 
c\, (a^2-b^2)\, (1-g^2)=0\,.\label{depicon}
\eea
Note that the full set of equations implies that (\ref{d6ans}) will be
Ricci-flat K\"ahler, since $d\ep_\3^R=0$ and $d\ep_\3^I=0$ are
together equivalent to $d\ep_\3=0$.  It is evident from the final
equation resulting from $d\ep_\3^I=0$, which is algebraic, that there
is a bifurcation into two non-singular possibilities, namely $a=b$ or
$g=1$ (other choice for the signs are equivalent, up to orientation).
The case $a=b$ gives rise to the first-order equations governing the
resolved conifold and a class of complex line bundles over $S^2\times
S^2$, whilst the case $g=1$ gives rise to the first-order equations
governing the deformed conifold.  The special case
\cite{berber,pagpop} of the complex line bundle over $S^2\times S^2$
with equal radii for the 2-spheres is a solution of both systems of
first-order equations.  For future purposes, where we shall be taking
these various Ricci-flat K\"ahler backgrounds as starting points for
linearised perturbative analyses, it is convenient to present the
first-order equations as follows:
\bea
&&\!\!\!\!\!\!\!\!\! \!\!\!\!\!\!\!\!\! 
\underline{a=b=A,\ c=C,\ g=G}:\nn\\
&&\nn\\
&& \dot A= -\, \fft{C}{4A}\,,\quad 
   \dot C = -1 +\fft{C^2\, (1+G^2)}{4A^2\, G^2}\,,\quad
   \dot G =-\, \fft{C\, (1-G^2)}{4 A^2\, G}\,,\label{resolvefo}\\
&&\nn\\
&&\!\!\!\!\!\!\!\!\! \!\!\!\!\!\!\!\!\! 
\underline{a=A,\ b=B,\ c=C,\ g=1}:\nn\\
&&\nn\\
&&\dot A =\fft{(A^2-B^2-C^2)}{4B\, C}\,,\quad 
\dot B =\fft{(B^2-A^2-C^2)}{4A\, C}\,,\quad 
\dot C =\fft{(C^2-A^2-B^2)}{2A\, B}\,.\label{deformfo}
\eea

\subsection{Linearised perturbations}\label{6linsec}

   We shall focus our attention on perturbations away from the
Ricci-flat K\"ahler conditions (\ref{resolvefo}).  The perturbed
metrics will be taken to lie within the same general class
(\ref{d6ans}), but no longer satisfying all three of the conditions
(\ref{djcon}), (\ref{deprcon}) and (\ref{depicon}) for $SU(3)$
holonomy.  We shall still require, however, that the perturbed metrics
be Ricci-flat.  In the light of our discussion in section \ref{fosec},
we shall also require that up to the linearised order 
in the perturbation the metrics be almost K\"ahler, and that
we can find some complexion angle $\gamma$ such that
$d \,\Re(e^{\im\, \gamma}\, \ep_\3)=0$ at the linearised order.

   By inspection of (\ref{djcon}) and (\ref{deprcon}), it is evident that
if we consider a perturbation taking the form
\be
a=A+\vep\, u\,,\qquad b=A - \vep\, u\,,\qquad c=C\,,\qquad g=G\,,
\label{abcpert}
\ee
where the upper-case functions satisfy the unperturbed Ricci-flat 
K\"ahler conditions (\ref{resolvefo}), then we shall have 
\be
dJ= 0+ O(\vep^2)\,,\qquad d\ep_\3^R = 0 + O(\vep^2)\,.
\ee
On the other hand, from (\ref{depicon}) we shall have $d\ep_\3^I= O(\vep)$.
Taking the complexion angle to be $\gamma=0$, we therefore satisfy all the 
requirements for being able to make a $G_2$ lift, provided that the 
perturbed metric is Ricci-flat and that $d\cF_\2=0$ and $d{*\cF_\2}=0$ 
at order $\vep$.
Substituting (\ref{abcpert}) into the Ricci tensor for (\ref{d6ans}),
we find that it vanishes at $O(\vep)$ if 
\be
\ddot u -\dot u \, \Big(\fft1{C} + \fft{C\, (1-G^2)}{4 A^2\, G^2}\Big) 
-u\, \Big(\fft1{C^2} - \fft1{2 A^2\, G^2} + \fft{3C^2\, 
       (1-2G^2)}{16 A^4\, G^4}\Big)
=0\,.\label{uddeq}
\ee

   By taking $\Psi_\3=\ep_\3^I$, we can solve for $\cF_\2$ from equation
(\ref{cfdef}), giving
\be
\cF_\2 = \vep\, \Big(-\fft{2\dot u}{A} 
  + \fft{2u}{A\, C} - \fft{u\, C}{2A^3}\Big)
\,(e^1\wedge e^2 + e^4\wedge e^5) - \fft{\vep\, u\, C\, 
(1-G^2)}{2A^3\, G^2}\, 
(2e^0\wedge e^3 -e^1\wedge e^5+ e^2\wedge e^4)\,.\label{cf2res}
\ee
Calculating $d\cF_\2$, we find that it vanishes at leading order in
$\vep$ if the condition (\ref{uddeq}) for Ricci-flatness is
satisfied.  We can also verify that $d{*\cF_\2}=0$ at this order.

   Using the first-order equations (\ref{resolvefo}) satisfied by
$A$, $C$ and $G$, we can find a first integral of (\ref{uddeq}), giving
\be
\dot u = \fft{u}{C} + \fft{u\, C\, (1-3G^2)}{4 A^2\, (1+G^2)\, G^2} +
\fft{k_2}{4A\, (1+G^2)}\,,\label{udeq}
\ee
where $k_2$ is an arbitrary integration constant.  This can be integrated
once more, again using (\ref{resolvefo}), to give the general solution
of (\ref{uddeq}),
\be
u= \fft{k_1\, (1+G^2)}{A\, C\, G^2} + \fft{A\, k_2}{4C\, G^2}\,,\label{ueq}
\ee
where $k_1$ is another constant of integration.

   Using (\ref{udeq}) and (\ref{ueq}), 
we can show that $\cF_\2$ can be written as
$\cF_\2= d(\a\, \Sigma_\3 + \beta\, \sigma_3)$, where the functions
$\a$ and $\beta$ are given by
\be
\a = -\fft{\vep\, k_2}{2G^2} - 
    \fft{2\vep\, k_1\, (1-G^2)}{A^2\, G^2}\,,\qquad
\beta = \fft{\vep\, k_2\,(1-2G^2) }{2G^2} +
 \fft{2\vep\, k_1\, (1-G^2)}{A^2\,G^2}\,.
\ee
This implies that $(\a\, \Sigma_3 + \beta\, \sigma_3) = 
\vep\, k_2\, (dz+ B_\1)$, 
where $\cA_\1=\vep\, k_2\, B_\1$ is the Kaluza-Klein vector and 
\bea
B_\1&=&-\, \fft{(1-G^2)\, (4k_1 + k_2\, A^2)}{k_2\, A^2\, G^2}\,  
dy  + \Big[ \fft{(1-2G^2) }{2G^2} +
 \fft{2k_1\, (1-G^2)}{k_2\, A^2\,G^2}\Big]
\, \cos\theta\, d\phi
\nn\\
&&- \Big[\fft{1}{2G^2} +
    \fft{2 k_1\, (1-G^2)}{k_2\, A^2\, G^2} \Big]\, \cos\td\theta\, d\td\phi
\,.
\eea
Here $y=\ft12(\psi-\wtd\psi)$, and 
neither $\cA_\1$ nor the six-dimensional metric (\ref{d6ans}) involves 
the coordinate $z=\ft12(\psi+\wtd\psi)$.  

  We are now in a position to lift the 
perturbed Ricci-flat metric to seven dimensions, using Kaluza-Klein
formalism discussed in section \ref{kksec}.  We therefore obtain
\bea
ds_7^2 &=& dt^2 + a^2\, [(\Sigma_1 + g\, \sigma_1)^2 +
                          (\Sigma_2 + g\, \sigma_2)^2]
+ b^2\,  [(\Sigma_1 - g\, \sigma_1)^2 +
                          (\Sigma_2 - g\, \sigma_2)^2]\nn\\
&&+ c^2\, (\Sigma_3-\sigma_3)^2 + \vep^2\, \lambda^2 \, (dz+\cA_\1)^2
\,,\nn\\
&=&  dt^2 + a^2\, [(\Sigma_1 + g\, \sigma_1)^2 +
                          (\Sigma_2 + g\, \sigma_2)^2]
+ b^2\,  [(\Sigma_1 - g\, \sigma_1)^2 +
                          (\Sigma_2 - g\, \sigma_2)^2]\nn\\
&&+ c^2\, (\Sigma_3-\sigma_3)^2 + f^2\, (\Sigma_3+ g_3\, \sigma_3)^2
\,,\label{d7met}
\eea
where $f=\a$ and $g_3=\beta/\a$.  Defining $e^6\equiv f\, 
 (\Sigma_3+ g_3\, \sigma_3)$, we see that the almost-K\"ahler 
structure $J=\Psi_2$ and the 3-form $\Psi_\3=\ep_\3^I$ in six dimensions
lift, using (\ref{3red}), to give
\bea
\hat \Phi_\3&=&  e^0\wedge ( e^1\wedge  e^4 + 
 e^2\wedge  e^5 +  e^3\wedge  e^6)
+( e^1\wedge  e^2- e^4\wedge  e^5)\wedge  e^3 \nn\\
&&+ ( e^1\wedge  e^5- e^2\wedge  e^4)\wedge  e^6
\,,\label{d73form}
\eea

   We have obtained a class of metrics in seven dimensions which, by
construction, have $G_2$ holonomy at the linearised order in $\vep$.
In fact, we have arrived at a linearised version of the general
description $G_2$ metrics with $S^3\times S^3$ principal orbits that
was obtained in \cite{munify}, where a system of five first-order
equations for the metric ansatz (\ref{d7met}) was obtained.

\subsection{Explicit discussion}\label{expldisc}

  The general solution of the first-order equations (\ref{resolvefo}) 
is given by
\be
A^2 = \ft12 (r+ \ell_1^2) \,,\qquad C^2= \fft{r\, (2r^2 + 
3(\ell_1^2 + \ell_2^2)\, r + 6\ell_1^2\, \ell_2^2)}{3(r+\ell_1^2)\,
(r+\ell_2^2)}\,,\qquad G^2 = \fft{r+\ell_2^2}{r+\ell_1^2}\,,
\label{gensol}
\ee
where $r$ is related to $t$ by $dr= C\, dt$, and $\ell_1$ and $\ell_2$
are constants.  The radial variable satisfies $r\ge 0$, where $r=0$ is
the bolt. If $\ell_1$ and $\ell_2$ are both non-zero the solution
corresponds to the one found in \cite{pando2} with an $S^2\times S^2$
bolt.  The special case $\ell_1=\ell_2$ is the metric found in
\cite{berber,pagpop}.  Regularity at $r=0$ requires that the
coordinate $y=(\psi-\wtd\psi)/2$ on the $U(1)$ fibre over $S^2\times
S^2$ have period $\pi$, and hence the principal orbits are
$T^{1,1}/Z_2$ \cite{pagpop,pando2}.  If $\ell_2=0$, the bolt is
instead an $S^2$, and the metric is the resolved conifold found in
\cite{candel}.  Regularity at $r=0$ now requires that $y$ have period
$2\pi$, and so the principal orbits are $T^{1,1}$.  (Taking $\ell_1=0$
instead gives the resolved conifold again, with the roles of the
$\Sigma_i$ and $\sigma_i$ interchanged.)

   The explicit expression for the perturbation function $u$ can be
obtained by substituting (\ref{gensol}) into (\ref{ueq}).  We are
interested in obtaining a perturbation which is regular for all $r$.
In particular, we see that to obtain regularity at $r=0$ we must
choose
\be
k_1 = -\, \fft{k_2\,\ell_1^4}{8(\ell_1^2+\ell_2^2)}\,.
\ee
The function $u$ is then given by
\be
u= \fft{\sqrt3\, k_2\, r^{1/2}\, [r\, (\ell_1^2+\ell_2^2) + 2\ell_1^2\, 
\ell_2^2]}{4(\ell_1^2 + \ell_2^2)\, \sqrt{r+\ell_2^2}\, \sqrt{
2r^2 + 3(\ell_1^2+\ell_2^2)\, r + 6\ell_1^2\, \ell_2^2}}\,.
\ee
It is easily seen that this is indeed regular for all $r$.  Thus we have
succeeded in obtaining a perturbation of the resolved conifold and of the
Ricci-flat K\"ahler metrics in \cite{berber,pagpop} and \cite{pando2}
that remains Ricci-flat at the linearised order, but which is no longer
K\"ahler (although it is still almost-K\"ahler).  The
resolved conifold corresponds to setting $\ell_2=0$, implying that
the original unperturbed metric, for which
\be
A^2= \ft12(r +\ell_1^2)\,,\qquad C^2 = 
\fft{r\, (2r+3\ell_1^2)}{3(r+\ell_1^2)}\,,\qquad 
G^2=\fft{r}{r+\ell_1^2}\,,
\ee
is perturbed according to (\ref{abcpert}) with
\be
u=\fft{\sqrt3 k_2\, r^{1/2}}{4\sqrt{2r+3 \ell_1^2}}\,.
\ee

   After lifting to seven dimensions the perturbed metrics give
regular metrics of $G_2$ holonomy.  The perturbed conifold gives a
linearisation around the Gromov-Hausdorff limit of the $\bD_7$
metrics found in \cite{mconifold}, while the perturbed metrics with
$\ell_1$ and $\ell_2$ both non-zero give linearisations around the
Gromov-Hausdorff limits of the $\wtd\bC_7$ metrics found in
\cite{munify}.  In particular, this provides an analytic
demonstration, to leading non-trivial order in perturbations, of the
conclusions reached by numerical analysis in \cite{mconifold} and
\cite{munify}, namely that the first-order equations for $G_2$
holonomy do indeed admit regular solutions of the types $\bD_7$ and
$\wtd\bC_7$.

   Evidence for another class of ALC $G_2$ metrics, denoted by
$\bB_7$, was found in \cite{cglpg2spin7,brgogugu}.  These have the
topology $\R^4\times S^3$, and for the limiting value of a non-trivial
parameter one obtains $S^1$ times the deformed conifold as the
Gromov-Hausdorff limit.  However, in this case the circle action whose
length stabilises at large distance collapses to zero length at short
distance, and so from a six-dimensional viewpoint the dilaton diverges
there.  This means that there is no analogous smooth Ricci-flat
perturbation of the deformed conifold.

   It is interesting to recall that after imposing the condition of
$SU(3)$ holonomy on the metric ansatz (\ref{d6ans}), we found a
bifurcation of the first-order equations into two branches, according
to whether the algebraic constraint in (\ref{depicon}) is solved by
taking $a=b$ or $g=1$.  The former includes the resolved conifold and
the metrics on the $\R^2$ bundles over $S^2\times S^2$, whilst the
latter includes the deformed conifold.  We obtained $G_2$ metrics
by considering perturbations of Calabi-Yau metrics in the first
branch, and these satisfy linearisations of 
first-order equations for $G_2$ holonomy in $D=7$. The full $G_2$
equations in $D=7$ in fact also encompass the 6-dimensional first-order
equations in the second branch, where $g=1$.  In other words the two
branches of first-order equations for $SU(3)$ holonomy, which are
disjoint in $D=6$, become unified in $D=7$ where they can be viewed as
two different Gromov-Hausdorff limits of the first-order equations for
$G_2$ holonomy obtained in \cite{munify}.  

\section{Almost $G_2$ metrics}

   Many of the ideas discussed above can also be applied to
eight-dimensional metrics of Spin(7) holonomy with a circle action.
From these considerations, we are led to introduce the notion of an
``Almost $G_2$ Metric.''

\subsection{Kaluza-Klein reduction of a Spin(7) metric}

   Given a Spin(7) metric of the form
\be
d\hat s_8^2 = e^{-\fft1{\sqrt{15}}\phi}\, ds_7^2 + e^{\fft{5}{\sqrt{15}}
\phi}\, (dz+\cA_\1)^2\,,
\ee
the calibrating self-dual 4-form $\hat\Phi_\4$ reduces to 
\be
\hat \Phi_\4 = \Phi_\3\wedge (dz+\cA_\1) + 
e^{-\fft{3}{\sqrt{15}}\phi}\,  {*\Phi_\3}
\ee
in seven dimensions, where $*$ denotes the seven-dimensional Hodge
dual.  The algebraic properties of $\hat \Phi_\4$ imply that $\Phi_3$ 
is $G_2$ invariant and, with the appropriate scaling, it 
satisfies the associativity property (\ref{octmult}).  
The closure condition $d\hat\Phi_\4=0$ that implies Spin(7)
holonomy therefore gives
\be
d\Phi_\3=0\,,\qquad d(e^{-\fft{3}{\sqrt{15}}\phi}\,
{*\Phi_\3})= \Phi_\3\wedge \cF_\2\,,
\ee
where $\cF_\2=d\cA_\1$.  
  
   If we consider a situation where $\cF_\2$ is small, of order $\vep$, then
$\phi$ will be a constant up to order $\vep$, leading to
\be
d\hat s_8^2 = ds_7^2 + (dz+\cA_\1)^2\,,\label{d8d7met}
\ee
where $ds_7^2$ is Ricci-flat up to linear order in $\vep$.  The 3-form
$\Phi_\3$ satisfies 
\be
d\Phi_\3=0\,,\qquad d{*\Phi_\3}= \Phi_\3\wedge \cF_\2\,,\label{almostg2}
\ee
at this linearised order.  

    It is useful to introduce the notion of an {\it almost $G_2$
metric}, namely a 7-metric admitting a $G_2$-invariant associative
3-form $\Phi_\3$ that is closed, but not necessarily co-closed.  It is
easy to show that if $X_\2$ is any 2-form, then $\Phi_\3\wedge X_\2=0$
implies $X_\2=0$, and hence one can solve for $\cF_\2$ uniquely from
(\ref{almostg2}).  Thus if a Ricci-flat 7-metric is almost $G_2$, and
if in addition it is such that $d\cF_\2=0$ and $d{*\cF_\2}=0$, then in
a linearisation around its $G_2$ holonomy limit it can be lifted to
give a Spin(7) metric.  It is natural to conjecture that there is an
analogue of Sekigawa's theorem \cite{sekigawa} for almost $G_2$
manifolds that are Ricci-flat; \ie if they are complete and
non-singular, they must be $G_2$ manifolds in the standard sense.
Nonetheless, as we shall see below, at the level of linearised
perturbations there can be complete and non-singular almost $G_2$
Ricci-flat deformations.  These remain complete and non-singular,
but cease to be Ricci-flat, beyond the linear level.

\subsection{Almost $G_2$ metrics with $SU(3)/(U(1)\times U(1))$ principal
orbits}

   In the notation of \cite{cglphyper}, we introduce a set of left-invariant
1-forms $(\sigma_1,\sigma_2,\Sigma_1,\Sigma_2,\nu_1,\nu_2,\lambda, Q)$
for $SU(3)$, which satisfy the exterior algebra
\bea
d\sigma_1 &=& -\ft12 \lambda\wedge \sigma_2 - \nu_1\wedge \Sigma_2 -
\nu_2\wedge \Sigma_1 - \ft32 Q\wedge \sigma_2\,,\nn\\
d\sigma_2 &=& \ft12 \lambda\wedge \sigma_1 + \nu_1\wedge \Sigma_1 -
\nu_2\wedge \Sigma_2 + \ft32 Q\wedge \sigma_1\,,\nn\\
d\Sigma_1 &=& \ft12 \lambda\wedge \Sigma_2 - \nu_1\wedge \sigma_2 +
\nu_2\wedge \sigma_1 - \ft32 Q\wedge \Sigma_2\,,\nn\\
d\Sigma_2 &=& -\ft12 \lambda\wedge \Sigma_1 + \nu_1\wedge \sigma_1 +
\nu_2\wedge \sigma_2 + \ft32 Q\wedge \Sigma_1\,,\nn\\
d\nu_1 &=& -\lambda\wedge \nu_2 - \sigma_2\wedge \Sigma_1 +
\sigma_1\wedge \Sigma_2\,,\nn\\
d\nu_2 &=& \lambda\wedge \nu_1 + \sigma_1\wedge \Sigma_1 +
\sigma_2\wedge \Sigma_2\,,\nn\\
d\lambda &=& 2\sigma_1\wedge \sigma_2 - 2 \Sigma_1\wedge \Sigma_2  
    + 4 \nu_1\wedge \nu_2\,,\nn\\
dQ  &=& 2\sigma_1\wedge \sigma_2 + 2\Sigma_1\wedge\Sigma_2\,.\label{d8ext}
\eea
Here $\lambda$ and $Q$ are the 1-forms along the $U(1)\times U(1)$ 
subalgebra.

   We consider 7-metrics of the form
\be
ds_7^2 = dt^2 + a^2\, (\sigma_1^2 +\sigma_2^2) + 
b^2\, (\Sigma_1^2 + \Sigma_2^2) + c^2\,(\nu_1^2 + \nu_2^2) \,.\label{d7calmet}
\ee
If we define the 3-form $\Phi_\3$ by
\bea
\Phi_\3 &=& e^0\wedge (e^1\wedge e^2 -e^3\wedge e^4 - e^5\wedge e^6)
  + e^1\wedge e^3\wedge e^5 - e^1\wedge e^4\wedge e^6 \nn\\
&& + e^2\wedge e^3\wedge e^6 + e^2\wedge e^4\wedge e^5\,,
\eea
where the vielbein is taken to be
\be
e^0=dt\,,\quad e^1 = a\, \sigma_1\,,\quad e^2 = a\, \sigma_2\,,\quad 
e^3 = b\, \Sigma_1\,,\quad e^4 = b\, \Sigma_2\,,\quad 
e^5 = c\, \nu_1\,,\quad e^6 = c\, \nu_2\,,\label{ortho}
\ee
then we find that closure and co-closure imply
\bea
d\Phi_\3=0:&& (a\, b\, c)^\cdot_{\phantom \Xi} = a^2+b^2+c^2\,,
\label{close}\\
d{*\Phi_\3}=0:&& (a^2\, b^2)^\cdot_{\phantom \Xi} =
                 (b^2\, c^2)^\cdot_{\phantom \Xi} =
                 (c^2\, a^2)^\cdot_{\phantom \Xi} = 4a\, b\, c\,.
  \label{coclose}
\eea
Together, these conditions produce the first-order equations for
$G_2$ holonomy\footnote{It happens in this particular case that the
equations following from $d{*\Phi_\3}=0$ imply that $d\, \Phi_\3=0$,
but this is not true in general.} that were obtained in \cite{cglpg2spin7};
\be
\dot a = \fft{b^2+c^2-a^2}{b\, c}\,,\quad
\dot b = \fft{c^2+a^2-b^2}{c\, a}\,,\quad
\dot c = \fft{a^2+b^2-c^2}{a\, b}\,.\label{flageq}
\ee
It was shown in \cite{cglpg2spin7} that these could be solved completely,
leading to $G_2$ metrics that are in general singular, except in the case
that any two of the three functions $a$, $b$ and $c$ are set equal.  Under
those circumstances, the solution then gives rise to the long-known AC
metric of $G_2$ holonomy on the $\R^3$ bundle over $\CP^2$, found in
\cite{brysal,gibpagpop}.  One could say, therefore, that perturbations
of the AC metric on the $\R^3$ bundle over $\CP^2$, within the class
described by (\ref{d7calmet}), will be singular if one requires that the
perturbed metric also have $G_2$ holonomy.

   Our goal now will be to consider perturbations, again contained within
the class (\ref{d7calmet}), which are Ricci-flat but no longer of $G_2$ 
holonomy.  In fact, we shall seek {\it almost $G_2$} perturbed
metrics, with $d\Phi_\3=0$ but $d{*\Phi_\3}\ne 0$. 
We consider perturbed metrics for which
\be
a=A+\vep\, u\,,\qquad b=A-\vep\, u\,,\quad c=C\,,
\ee
where $A$ and $C$ satisfy the first-order equations
\be
\dot A = \fft{C}{A}\,,\qquad \dot C= 2 -\fft{C^2}{A^2}\label{r3cp2eq}
\ee
that follow from setting $a=b=A$, $c=C$ in (\ref{flageq}).  We then find that
the Ricci tensor of the metric (\ref{d7calmet}) will vanish at order $\vep$ 
if the perturbing function $u$ satisfies the second-order equation
\be
\ddot u + \fft{\dot u}{C} + \Big( \fft6{A^2} + \fft{2C^2}{A^4} 
- \fft{8}{C^2}\Big)\, u=0\,.
\ee
Using (\ref{r3cp2eq}) we can integrate this once, to give
\be
(A\, C^2\, u)^\cdot_{\phantom \Xi} = k_2\, C^2\,,\label{dueqg2}
\ee
where $k_2$ is an arbitrary constant of integration.  There is no easy way
to integrate this again abstractly, unless one sets $k_2=0$.  Making this
choice implies that the equations (\ref{flageq}) for $G_2$ holonomy 
are satisfied.  Since, as was shown in \cite{cglpg2spin7}, the associated
solutions will necessarily be singular when $u\ne0$, we therefore wish to
keep $k_2$ non-zero here.\footnote{We have explicitly verified that if
$k_2$ is non-zero, then the integrability condition $R_{abcd}\,
\Gamma^{cd}$ has no zero eigenvalues.  This proves that the perturbed
metric does not have $G_2$ holonomy, and thereby excludes the
possibility that there might have existed a perturbed associative
3-form that was still co-closed as well as closed.}

   The general solution to the unperturbed equations (\ref{r3cp2eq}) can be
written, after eliminating trivial integration constants, as
\be
A^2 = r^2\,,\qquad C^2 = r^2\, (1-r^{-4})\,,
\ee
where $r$ is a new radial variable defined in terms of $t$ by $dr=
(1-r^{-4})^{1/2}\, dt$.  This gives the AC metric on the $\R^3$ bundle 
over $\CP^2$ that was found in \cite{brysal,gibpagpop}, which has a 
$\CP^2$ bolt at $r=1$.  Substituting
into (\ref{dueqg2}), we can solve for $u$, giving
\be
u= \fft{k_1\, r}{r^4-1} + \fft{k_2\, r^2}{3\sqrt{r^4-1}} 
+ \fft{2k_2\, r\, F(\arcsin r^{-1}| -1)}{3(r^4-1)}\,,
\ee
where $F(\phi|m)$ is the elliptic integral of the first kind.  The function
$u$ is regular at the bolt at $r=1$ if the constants are chosen so that
\be
k_1 = - \ft23 k_2\, K(-1)\,,
\ee
where $K(m)$ is the complete elliptic integral of the first kind.  We
find that $u$ is regular everywhere, with $u\sim \ft13 k_2\,
\sqrt{r-1}$ at short distance, and $u\sim \ft13 k_2$ at large distance.
Thus we have a regular Ricci-flat perturbation of the AC metric on the
$\R^3$ bundle over $\CP^2$.

    Up to linear order in $\vep$, we find that
\be
d\Phi_\3 = 0 + O(\vep^2)\,,\qquad d{*\Phi_\3}=  \Phi_\3\wedge \cF_\2\,,
\ee
which, after using using (\ref{dueqg2}), implies $\cF_\2 = 2\vep\,
k_2\, (\sigma_1\wedge \sigma_2 + \Sigma_1\wedge \Sigma_2)$.  From
 (\ref{d8ext}) we therefore have $\cF_\2=d\cA_\1$ with
\be
dz+ \cA_\1 = \vep\, k_2\, Q\,.
\ee
It is easily verified that $d{*\cF_\2}=0$, and hence all the requirements
for our perturbed 7-metric to admit a lift to an eight-dimensional
metric of Spin(7) holonomy are fulfilled.   From (\ref{d8d7met}), 
we obtain the Spin(7) metric
\be
d\hat s_8^2 = dt^2 + a^2\, (\sigma_1^2 +\sigma_2^2) + 
b^2\, (\Sigma_1^2 + \Sigma_2^2) + c^2\,(\nu_1^2 + \nu_2^2) 
+ f^2\, Q^2 \,,\label{d8ans}
\ee
where $f$ is a small constant, given by $f=\vep\, k_2$.  The self-dual 
calibrating 4-form is given by
\be
\hat\Phi_\4 = \Phi_\3\wedge e^7 + {*\Phi_\3}\,,
\ee
where $e^7\equiv f\, Q$.

   This system in eight-dimensions can in fact be ``non-linearised.''
The metric (\ref{d8ans}) is precisely one that was considered in 
\cite{cglpg2spin7}, for Spin(7) metrics whose principle orbits are
the $N(k,\ell)=SU(3)/U(1)_{k,\ell}$ space with $k=1$, $\ell=-1$.  
The first-order equations implying Spin(7) holonomy in this case are
\bea
\dot a &=& \fft{b^2+c^2-a^2}{b\, c} - \fft{f}{a}\,,\nn\\
\dot b &=& \fft{a^2+c^2-b^2}{c\, a} + \fft{f}{b}\,,\nn\\
\dot c &=& \fft{a^2+b^2-c^2}{a\, b }\,,\nn\\
\dot f &=&  \fft{f^2}{a^2} - \fft{f^2}{b^2}\,.
\label{d8fo}
\eea
It was
shown in \cite{cglpg2spin7}, by performing a Taylor 
expansion, that these equations admit regular short-distance solutions in
which just the function $c$ goes to zero at $t=0$.  This corresponds to
an $S^5$ bolt.  By numerical analysis, it was also 
found in \cite{cglpg2spin7} that the metrics are regular at large distance,
provided that the initial values for the non-vanishing metric functions
on the bolt are chosen appropriately.  The metrics are in general ALC, 
with $f$ stabilising to a constant radius at infinity.  
Note that the radius $f$ of the circle remains finite and non-zero 
everywhere.  A non-trivial
parameter adjusts the value of this radius, while keeping the scale-size
of the bolt non-zero, and so we can take a Gromov-Hausdorff limit.  In
this limit, we get the product of a circle and the AC metric on the 
$\R^3$ bundle over $\CP^2$.  

   The perturbative analysis that we performed above has yielded the
metrics in seven-dimensions that correspond to the Kaluza-Klein reduction 
of the regular ALC metrics found in \cite{cglpg2spin7}, as one moves 
away from the Gromov-Hausdorff limit.

\section{Almost-K\"ahler Ricci-flat metrics}

   In the section \ref{d6kah}, we obtained examples, as linearised
perturbations of Calabi-Yau metrics, of six-dimensional metrics that are 
Ricci-flat and almost-K\"ahler but not K\"ahler.  In this section we 
obtain some analogous examples in higher dimensions.  

\subsection{Line bundles over $SO(n+2)/(SO(n)\times SO(2))$}

   Cohomogeneity one metrics with $SO(n+2)/SO(n)$ principal orbits
were discussed in \cite{cglpsten}.  
    Let $L_{AB}$ be the left-invariant 1-forms on the group manifold
$SO(n+2)$.  These satisfy
\be
dL_{AB} = L_{AC} \wedge L_{CB}\,.
\ee
We consider the $SO(n)$ subgroup, by splitting the index as
$A=(1,2,i)$.  The $L_{ij}$ are the left-invariant 1-forms for the
$SO(n)$ subgroup.  We make the following definitions:
\be
\sigma_i \equiv L_{1i}\,,\qquad \td\sigma_i \equiv L_{2i}\,,\qquad
\nu \equiv L_{12}\,.
\ee
These are the 1-forms in the coset $SO(n+2)/SO(n)$.  We have
\bea
&&d\sigma_i = \nu\wedge \td\sigma_i + L_{ij}\wedge \sigma_j\,,\quad
d\td\sigma_i = -\nu\wedge \sigma_i + L_{ij}\wedge
\td\sigma_j\,,\quad
d\nu = -\sigma_i\wedge \td\sigma_i\,,\nn\\
&&dL_{ij} = L_{ik}\wedge L_{kj} -\sigma_i\wedge \sigma_j -
\td\sigma_i\wedge \td\sigma_j\,.\label{exd}
\eea
Note that the 1-forms $L_{ij}$ lie outside the coset, and so one finds
that they do not appear eventually in the expressions for the
curvature (see also \cite{danstr,danstr2}).  The metrics were written as 
\be
ds^2 = dt^2 + a^2 \sigma_i^2 + b^2\, \td\sigma_i^2 + c^2\, \nu^2\,,
\label{stenans}
\ee
where $a$, $b$ and $c$ are functions of the radial coordinate $t$.

   The first-order equations that imply $SU(n+1)$ holonomy are 
\cite{cglpsten}
\be
\dot a = \fft{b^2+c^2-a^2}{2 b\, c}\,,\quad
\dot b = \fft{a^2+c^2-b^2}{2 a\, c}\,,\quad
\dot c = \fft{n\, (a^2+b^2-c^2)}{2 a\, b}\,.\label{stenfo}
\ee
Here, we shall consider perturbations of the metric (\ref{stenans})
around the special case $a=b=A$, $c=C$, for which we have
\be
\dot A= \fft{C}{2A}\,,\qquad \dot C = n -\fft{n\, C^2}{2A^2}\,,
\label{truncfo}
\ee
where
\be
a=A + \vep\, u\,,\qquad b=A-\vep\, u\,,\qquad c=C\,.\label{stenab}
\ee
Note that the perturbations we shall consider will be Ricci-flat, but they
will not be required to satisfy the first-order equations (\ref{stenfo}). 

   Requiring Ricci-flatness up to linear order in $\vep$, we find that
the perturbation $u$ must satisfy the second-order equation
\be
\ddot u + \Big(\fft{n}{C} + \fft{(n-2)\, C}{2 A^2}\Big)\, \dot u
+ \Big( \fft{n}{A^2} - \fft4{C^2} + \fft{3C^2}{4A^4}\Big)\, u=0\,.
\label{secondu}
\ee
Using (\ref{truncfo}), we can find a first integral of this, namely
\be
(u\, C^{2/n}\, A)^{\bf\cdot}_{\phantom \Xi} = 
\fft{\lambda\, C^{4/n-1}}{A^{2n-4}}\,,\label{firstu}
\ee
where $\lambda$ is an arbitrary constant.  If $\lambda=0$, this
first-order equation follows by substituting (\ref{stenab}) into the
first-order equations (\ref{stenfo}) for $SU(n+1)$ holonomy, but if
$\lambda\ne 0$ the perturbed Ricci-flat metric is not K\"ahler.
Unlike the previous example in (\ref{6linsec}), here the equation
(\ref{firstu}) cannot in general be easily integrated again
abstractly, and so we shall proceed to the explicit solution.

   The general solution of the first-order system (\ref{truncfo}) can be
written as
\be
A^2 = \fft{n\, r^2}{2(n+1)}\,,\qquad C^2 = \fft{n^2\, r^2}{(n+1)^2}\, 
\Big(1-\fft1{r^{2n+2}}\Big)\,,
\ee
in terms of a new radial coordinate $r$ defined by $dr= (1-
r^{-2n-2})\, dt$.  In fact the metric corresponding to this solution is
precisely in the class constructed in \cite{berber,pagpop}, for the special
case of the complex line bundle over the Einstein-K\"ahler metric on the
Grassman manifold $SO(n+2)/(SO(n)\times SO(2))$.  The radial coordinate
satisfies $r\ge 1$.  The general solution
for the perturbation $u$ is then given by
\be
u = r^{-2/n-1}\, \Big(k_1
 + k_2\, r^{4/n-2n+4}\, 
_2F_1[a_1, a_2;a_3; r^{-2/(n+1)}]\Big)\, 
\Big(1-\fft{1}{r^{2n+2}}\Big)^{-1/n}\,,\nn
\ee
where
\be
a_1 = \fft{n^2-2n-2}{n\, (n+1)}\,,\qquad a_2 = \fft{n-2}{n}\,,\qquad 
a_3= \fft{2n^2-n-2}{n\, (n+1)}\,.
\ee
Regularity of $u$ on the bolt at $r=1$ requires that the 
constants $k_1$ and $k_2$ be chosen such that
\be
k_1 = -\, \fft{k_2\, \Gamma[(n-1)/n]\, \Gamma[(2n^2-n-2)/(n\, (n+1))]}{
\Gamma[n/(n+1) ]}\,.
\ee
This implies that we have $u\sim (r-1)^{1/n}$ near to $r=1$.  At large
distance we then have
\be
u\sim \cases{r^3\,, & $n=1$\cr
             1\,, & $n=2$\cr
          r^{-2/n-1}\,, & $n\ge 3$\cr}
\ee
Thus we see that if $n\ge2$, the perturbation will be regular for all $r$.    
   
\section*{Acknowledgements}

   We are grateful to Vestislav Apostolov for helpful discussions.  G.W.G.,
H.L. and C.N.P. are grateful to UPenn for support and hospitality
during the course of this work.  

\section*{Note Added}

   After this work was completed, a paper appeared \cite{chisal} that overlaps
with some of our discussion of fibred $G_2$ manifolds.

\end{document}